%% file: main.tex
\documentclass[sigconf,screen]{acmart}
\setcopyright{acmlicensed}
\acmPrice{15.00}
\acmDOI{10.1145/3611643.3613076}
\acmYear{2023}
\copyrightyear{2023}
\acmSubmissionID{fse23ivr-p5-p}
\acmISBN{979-8-4007-0327-0/23/12}
\acmConference[ESEC/FSE '23]{Proceedings of the 31st ACM Joint European Software Engineering Conference and Symposium on the Foundations of Software Engineering}{December 3--9, 2023}{San Francisco, CA, USA}
\acmBooktitle{Proceedings of the 31st ACM Joint European Software Engineering Conference and Symposium on the Foundations of Software Engineering (ESEC/FSE '23), December 3--9, 2023, San Francisco, CA, USA}
\received{2023-05-03}
\received[accepted]{2023-07-19}

\usepackage{mystyle}

\hypersetup{
	pdftitle={Towards Top-Down Automated Development in Limited Scopes: A Neuro-Symbolic Framework from Expressibles to Executables},
	pdfauthor={Jian Gu, Harald C. Gall},
	pdfkeywords={code generation, deep learning, frame semantics, knowledge graph, requirement elicitation, program representation, software analytics}
}

\input{acronyms}

\begin{document}

\input{front/title}
\input{front/authors}
\input{front/abstract}
\input{front/keywords}
\input{front/copyright}
\maketitle

\input{contents/1.introduction}
\input{contents/2.approach}
\input{contents/3.roadmap}
\input{contents/4.related_work}
\input{contents/5.conclusions}

\clearpage
% \bibliography{references}
\onecolumn
\begin{multicols}{2}
\bibliography{references}
\end{multicols}

\end{document}

%% file: acronyms.tex
\DeclareAcronym{api}{
	short = API,
	long = {Application Program Interface}
}

\DeclareAcronym{awgn}{
	short = AWGN,
	long = {additive white Gaussian noise}
}

\DeclareAcronym{vae}{
	short = VAE,
	long = {Variational AutoEncoder}
}

\DeclareAcronym{bert}{
	short = BERT,
	long = {Bidirectional Encoder Representations from Transformers}
}

\DeclareAcronym{roberta}{
	short = RoBERTa,
	long = {Robustly optimized BERT approach}
}

\DeclareAcronym{ast}{
	short = AST,
	long = {Abstract Syntax Tree}
}

\DeclareAcronym{bpe}{
	short = BPE,
	long = {Byte-Pair Encoding}
}

\DeclareAcronym{cfg}{
	short = CFG,
	long = {Control Flow Graph}
}

\DeclareAcronym{dcg}{
	short = DCG,
	long = {Discounted Cumulative Gain}
}

\DeclareAcronym{gpt}{
	short = GPT,
	long = {Generative Pretrained Transformer}
}

\DeclareAcronym{ir}{
	short = IR,
	long = {Information Retrieval}
}

\DeclareAcronym{lstm}{
	short = LSTM,
	long = {Long Short-Term Memory}
}

\DeclareAcronym{clm}{
	short = CLM,
	long = {Casual Language Modeling}
}

\DeclareAcronym{mlm}{
	short = MLM,
	long = {Masked Language Modeling}
}

\DeclareAcronym{mem}{
	short = MEM,
	long = {Multimodal Embedding Model}
}

\DeclareAcronym{cp}{
	short = CP,
	long = {Continuous Pretraining}
}

\DeclareAcronym{if}{
	short = IF,
	long = {Intermediate Finetuning}
}

\DeclareAcronym{mmpf}{
	short = MMPF,
	long = {Massive Multitask Pre-Finetuning}
}

\DeclareAcronym{aif}{
	short = AIF,
	long = {Adaptive Intermediate Finetuning}
}

\DeclareAcronym{mrr}{
	short = MRR,
	long = {Mean Reciprocal Rank}
}

\DeclareAcronym{ndcg}{
	short = NDCG,
	long = {Normalized Discounted Cumulative Gain}
}

\DeclareAcronym{nlp}{
	short = NLP,
	long = {Natural Language Processing}
}

\DeclareAcronym{nlp_pt}{
	short = NLP\textsubscript{PT},
	long = {Next Line Prediction}
}

\DeclareAcronym{nmt}{
	short = NMT,
	long = {Neural Machine Translation}
}

\DeclareAcronym{nsp}{
	short = NSP,
	long = {Next Sentence Prediction}
}

\DeclareAcronym{rnn}{
	short = RNN,
	long = {Recurrent Neural Network}
}

\DeclareAcronym{cnn}{
	short = CNN,
	long = {Convolutional Neural Network}
}

\DeclareAcronym{tf-idf}{
	short = tf-idf,
	long = {term frequency–-inverse document frequency}
}

\DeclareAcronym{anova}{
	short = ANOVA,
	long = {ANalysis Of VAriance}
}

\DeclareAcronym{da}{
	short = DA,
	long = {Domain-Adaptive}
}

\DeclareAcronym{ta}{
	short = TA,
	long = {Task-Adaptive}
}

\DeclareAcronym{ma}{
	short = MA,
	long = {Multiphase Adaptive}
}

\DeclareAcronym{ca}{
	short = CA,
	long = {Concept Annotation}
}

\DeclareAcronym{ce}{
	short = CE,
	long = {Concept Extrapolation}
}

\DeclareAcronym{ci}{
	short = CI,
	long = {Concept Interpolation}
}

\DeclareAcronym{gru}{
	short = GRU,
	long = {Gated Recurrent Unit}
}

\DeclareAcronym{sota}{
	short = SOTA,
	long = {state-of-the-art}
}

\DeclareAcronym{lcs}{
	short = LCS,
	long = {Longest Common Sequences}
}

%% file: front/title.tex
\title[Towards Top-Down Automated Development in Limited Scopes \ldots]{Towards Top-Down Automated Development in Limited Scopes:\\A Neuro-Symbolic Framework from Expressibles to Executables}

%% file: front/authors.tex
\author{Jian Gu}
\affiliation{%
  \institution{Monash University}
  \city{Melbourne}
  \country{Australia}
}
\email{jian.gu@monash.edu}

\author{Harald C. Gall}
\affiliation{%
  \institution{University of Zurich}
  \city{Zurich}
  \country{Switzerland}
}
\email{gall@ifi.uzh.ch}

%% file: front/abstract.tex
\begin{abstract}
    Deep code generation is a topic of deep learning for software engineering (DL4SE), which adopts neural models to generate code for the intended functions. Since end-to-end neural methods lack domain knowledge and software hierarchy awareness, they tend to perform poorly w.r.t project-level tasks.
    To systematically explore the potential improvements of code generation, we let it participate in the whole top-down development from \emph{expressibles} to \emph{executables}, which is possible in limited scopes. In the process, it benefits from massive samples, features, and knowledge.
    As the foundation, we suggest building a taxonomy on code data, namely code taxonomy, leveraging the categorization of code information.
    Moreover, we introduce a three-layer semantic pyramid (SP) to associate text data and code data. It identifies the information of different abstraction levels, and thus introduces the domain knowledge on development and reveals the hierarchy of software.
    Furthermore, we propose a semantic pyramid framework (SPF) as the approach, focusing on software of high modularity and low complexity. SPF divides the code generation process into stages and reserves spots for potential interactions.
    In addition, we conceived preliminary applications in software development to confirm the neuro-symbolic framework.
    
    % My other main criticism is that despite sections 2.1 and 2.2, it is not very clear to me how a concrete implementation would work.
    % As in: it's easy to create a taxonomy, but how do you actually create an instance with data? Of course, code can be parsed, APIs can be analyzed, but the semantic bridging seems very hard to automate. I would like to see a more concrete step-by-step outline to go from conception to implementation. I would maybe shorten 2.1 and 2.2, or rather, merge them with section 2, where the concepts are already being discussed, and instead add a roadmap for reaching the goals because it's not clear to me how that could happen.
\end{abstract}

%% file: front/keywords.tex
\keywords{code generation, deep learning, frame semantics, knowledge graph, requirement elicitation, program representation, software analytics}

%% file: front/copyright.tex
% \acmConference[ASE 2022]{The 37th IEEE/ACM International Conference on Automated Software Engineering}{October 10-14, 2022}{Ann Arbor, MI, USA}

%% file: contents/1.introduction.tex
\section{Introduction}
\label{sec:introduction}

To promote the productivity of software development, using deep learning to assist related activities is a lasting research trend~\cite{Le2020DeepLF,Watson2020ASL}. For example, neural models trained on massive code-related data could assist programming through code generation~\cite{Lee2021TowardCG}.
However, we identified a few restrictions of neural methods waiting to be overcome.
First, models lack the necessary domain knowledge on software development. For example, models can merely learn the idiomatic usage of the relevant API dependency but cannot ensure validity or avoid possible misuse issues.
Second, models require accurate and complete functional descriptions, even related to the context and method call. The truth is that it is better to let models infer the requirement based on the software hierarchy.
Third, neural methods are weak in transparency and flexibility, thereby tend to underperform in scenarios where potential interactions are required, such as manual inspection or intervention.
Based on the above considerations, we propose adopting neural models in the whole top-down development from expressibles to executables. In this way, we explore performance improvements in various scenes, by providing domain knowledge and software hierarchy to code generation.
Considering the difficulty, the research scope is limited to softwares of high modularity and low complexity.

\begin{figure}[htbp]
    \centering
    \includegraphics[width=0.8\linewidth]{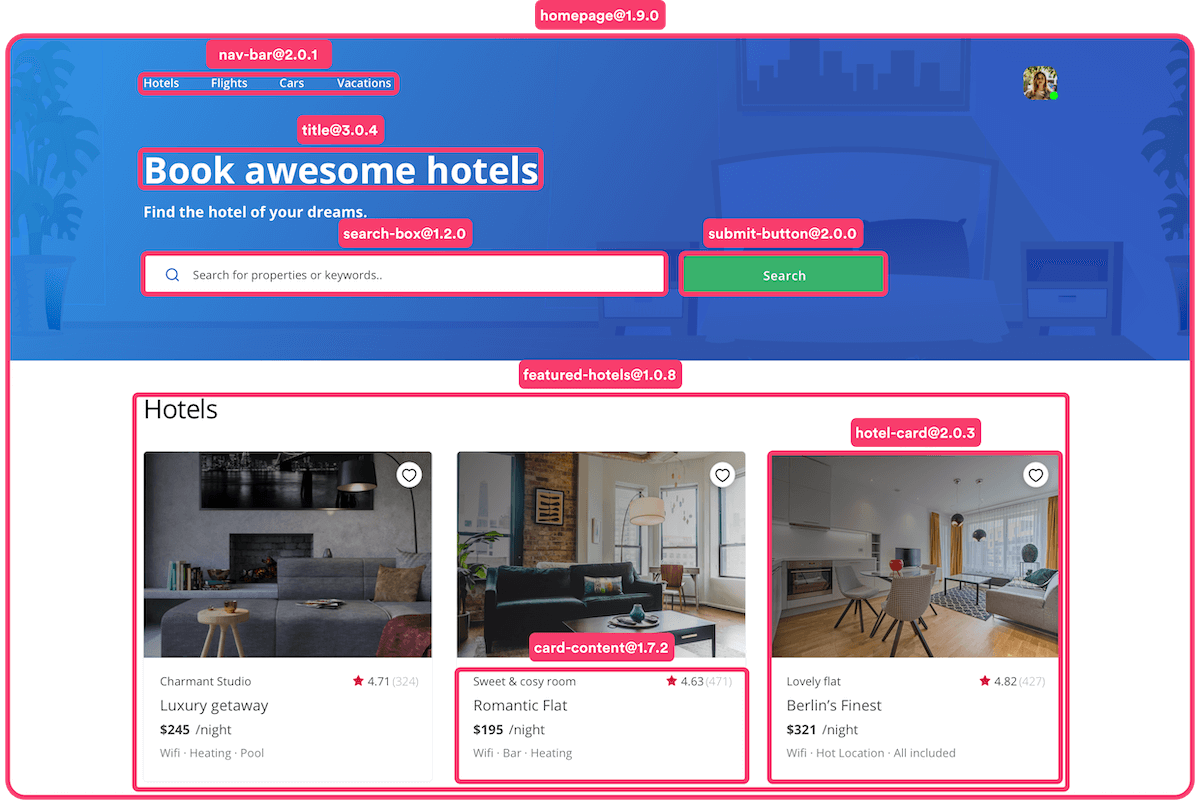}
    \caption{\small Diagram of modular WebUI development, © Bit.Cloud
    \protect\cite{web_bitcloud_20220520}, where each UI component in the webpage is supported by a separate dependency, and meanwhile filled with specific resources or values.}
    \label{fig:modular}
\end{figure}

The webpage in \cref{fig:modular} is an illustrative example of top-down code generation. Its implementation from expressibles to executables can be a 3-step process: (1) everything starts with a prototype, which clearly describes the webpage layout. For example, there should be a navigation bar, an avatar and a search bar in the header, and a grid view to display cards in the webpage body; and (2) a knowledge engine that knows available libraries suggest API usages to fulfill the requirements, such as recommending ``featured-hotels@1.x'' and ``hotel-card@2.x'' for the grid view and cards, with the compatibility consideration; and (3) a set of utility codes accessing algorithms and data resources arrange and control UI components, including personalizing the product list, loading the product detail.

To achieve the goal of adopting code generation in the top-down development process, we start with taxonomy on code data, namely code taxonomy, in the light of code information.
On the basis of code taxonomy, we designed the scheme of a three-layer semantic pyramid (SP) to associate program realizations with human intentions. In each SP layer, we take corresponding techniques to organize data in the form of graphs.
By taking SP as the mental model, we plan to further construct a semantic pyramid framework (SPF). SPF makes code data high-quality and well-organized, and besides, introduces program semantics and domain knowledge.

% Consequently, the significance of our approach is as follows:
% \begin{itemize}
%     \item a taxonomy on code data based on code information.
%     \item a mental model to associate intentions and realizations.
%     \item a framework incorporating feature bases, knowledge bases, and sample bases to support top-down code generation.
% \end{itemize}

%% file: contents/2.approach.tex
\section{Approach}
\label{sec:approach}

\paragraph{\textbf{Code Taxonomy}} We define the term ``code taxonomy'' as the practice of categorization of code data.
It is straightforward to build a taxonomy referring to software purposes or topics, however, that manner seems insufficient in reflecting the capability of methods, especially for code generation.
Therefore, we describe the task from another viewpoint: \emph{beyond the influence of software purpose or topic, by what basis could we further build taxonomy on code data?}

In our approach, we regard code data as a mixture of different types of information and disentangle this information to build code taxonomy. Considering the inherent properties of code data and the semantic associations with related text data, we classify code information into 3 groups and 6 types, as shown in \cref{tab:information}. Based on the categorization of code information, we suggest quantifying the \emph{information abundance} of data samples on each information type for various usages. For example, when we evaluate code generation respecting the difficulty of data samples, it is natural to assess the results separately for cases like whether there are API dependencies.

\begin{table}[!tb]
    \caption{Categorization of code information}
    \label{tab:information}
    \centering
    \resizebox{\linewidth}{!}{%
        \input{tables/information}
    }
\end{table}

\begin{figure}[htbp]
    \centering
    \includegraphics[width=0.8\linewidth]{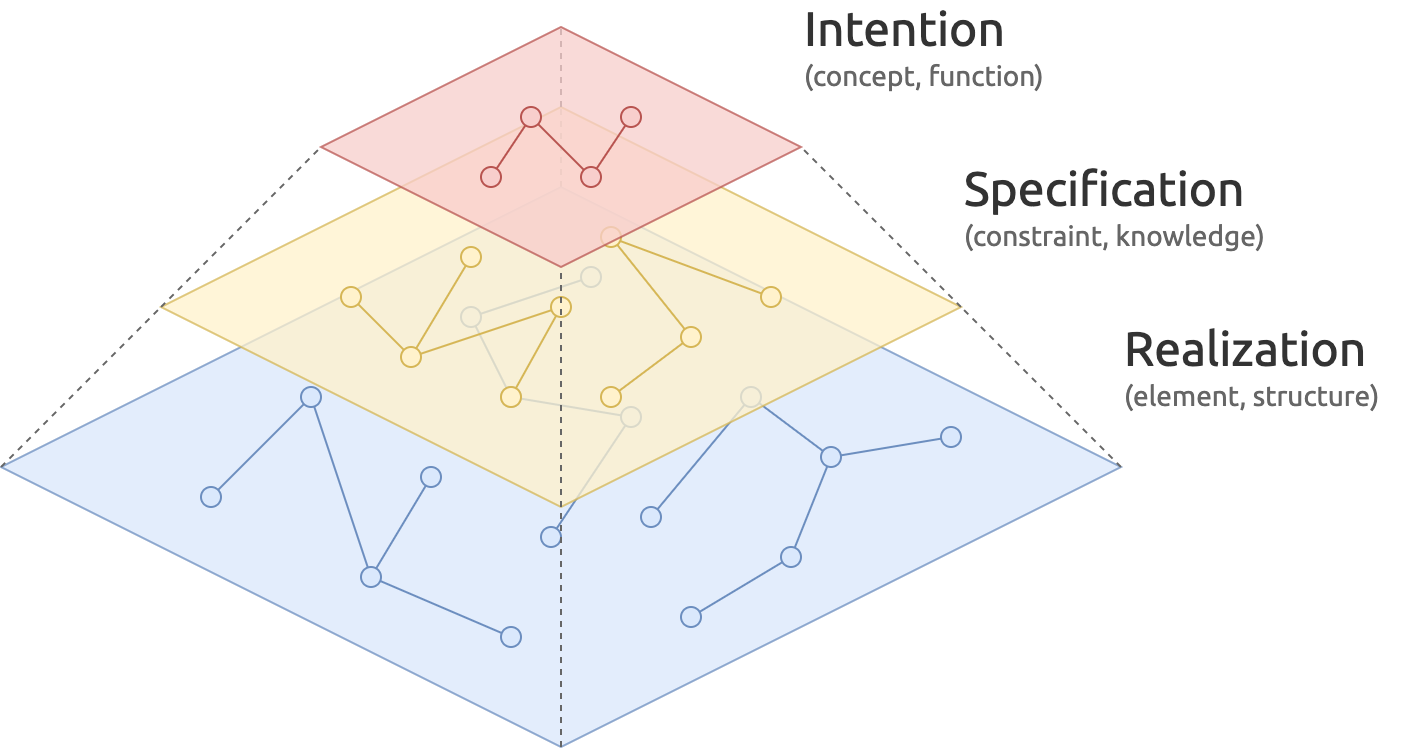}
    \caption{\small Diagram of the three-layer semantic pyramid, where the code information of each layer is specified. The higher the layer, the closer to text data, and vice versa, the closer to code data.}
    \label{fig:pyramid}
\end{figure}

\paragraph{\textbf{Semantic Pyramid}}
We propose a semantic-aware framework as the follow-up work of introducing code taxonomy for code generation. SPF guarantees the recognizability of code data via \emph{code assetization} and establishes the connectivity between intentions and realizations via \emph{semantic bridging}. Its mental model is visualized as a three-layer semantic pyramid, shown in \cref{fig:pyramid}.

The code information is divided into 3 abstraction layers. The intention layer cares about semantic information, namely semantic concepts and logic functions of code data. The specification layer is about the outline of code implementations, such as constraints and dependencies that cannot be ignored. The realization layer is about the detail of code implementations, where concrete elements or structures matter.
Every node of any graph in the upper layer points to the complete graph in the lower layer, and each graph in the lower layer is pointed to by multiple nodes in the upper layer.

The involved information is intended to be organized in the form of graphs even though it corresponds to different abstraction levels in different layers.
The reason is that graph-form expressions are ideal as the medium to explicitly represent various information, and performant to establish connections with each other. Considering the differences of the information itself and their roles in the framework, we introduce suitable techniques to build graphs and find ways to bridge different graphs, and thus associate intentions to realizations crossing the concept hierarchy of software.

\begin{table}[!tb]
    \caption{Definitions of the datatype for code taxonomy}
    \label{tab:taxonomy}
    \centering
    \resizebox{\linewidth}{!}{%
        \input{tables/taxonomy}
    }
\end{table}

\paragraph{\textbf{Semantic Pyramid Framework}}
To describe the mechanism of our framework, we distinguish the involved datatypes based on their information abundance following the code taxonomy. For simplicity, we prefer the qualitative division but not the quantitative one. As shown in \cref{tab:taxonomy}, we use the check mark (\cmark) to indicate a datatype is rich in some types of code information, and the cross mark (\xmark) for the opposite case. If an information type is not that important to a datatype, we leave blank the marked area.

\begin{figure}[htbp]
    \centering
    \includegraphics[width=0.8\linewidth]{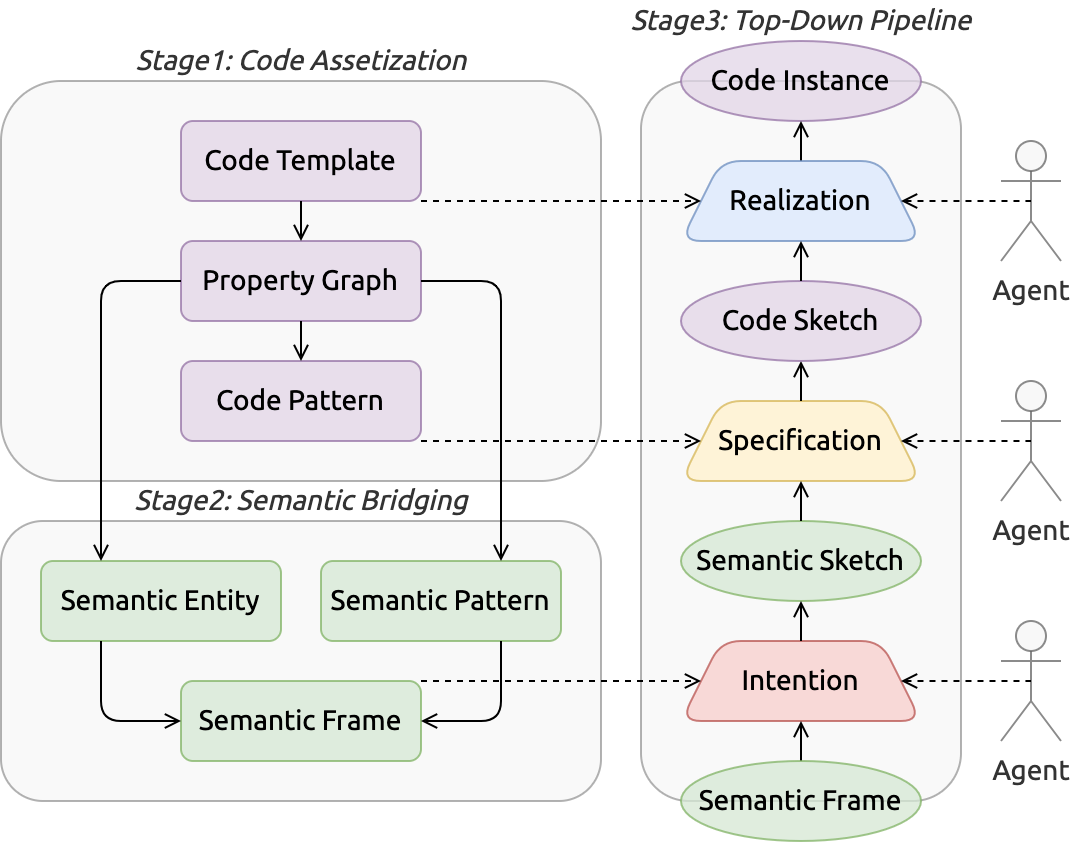}
    \caption{\small Overview of the semantic pyramid framework, where rectangles are databases of the corresponding datatype, trapezoids are spots for human-machine interactions, and ellipses represent data instances. The color is to distinguish datatypes and procedures.}
    \label{fig:overview}
\end{figure}

By characterizing the role of datatypes defined following code taxonomy, we divide the mechanism of the framework into three stages, as illustrated in \cref{fig:overview}.
In the first stage named \emph{code assetization}, code data is first processed into subprograms as templates to increase the sample richness. Then, we use a property graph as the code representation for the higher recognizability of samples \cite{Yamaguchi2014ModelingAD}. Eventually, we mine code patterns from code property graphs to reduce the effect of variants.
In the second stage named \emph{semantic bridging}, leveraging the informative property graphs, we extract semantic entities and patterns to collect involved concepts and functions. Furthermore, we compose semantic frames based on concepts and functions with external knowledge bases.
In the third stage named \emph{top-down pipeline}, we introduce code generation to assist top-down development, leveraging the databases on features, knowledge, and samples from the former stages. It divides the development into stages and reserves spots for interactions, corresponding to the abstraction layers in the semantic pyramid.

\subsection{Code Assetization}

We define \emph{Code Assetization} as organizing code data into samples, following code taxonomy. It focuses on the realization layer.

In the process of code assetization, SPF decouples semantics, syntactics, and attributes information from code samples. We need to generate high-quality representations of code samples~\cite{Whang2021DataCA} to support operations such as retrieval and clustering. As code data could be parsed as ASTs, CFGs, and PDGs, it is a cheap but precise practice to first convert code data into the graph form, as the joint data structure of the above-mentioned parsed results~\cite{Yamaguchi2014ModelingAD}.

We build a sample base of code templates and code patterns through data processing. Except for the handling of the elements, the code template is roughly equivalent to the code sample, since both reserve the divergence of concrete implementations. In contrast, the code pattern represents a bunch of similar code samples and has no tolerance for variants. Leveraging diff computation~\cite{Falleri2014FinegrainedAA} and subprogram division~\cite{Alet2021ALB}, code patterns are extracted from code templates but shorter in length and fewer in amount. As the property graph is preferred to represent code data, code patterns could be seen as the idiomatic usage of a partial code template.

The realization layer is focused on the synergy of code patterns with extra knowledge bases. Since the divergence in code samples is mainly caused by contexts and dependencies, we thus define two types of knowledge bases to complement the information missed by code patterns. One knowledge base is on internal context, which determines the constraints of code templates. The other one is on external dependencies, which determines the API dependencies.

\subsection{Semantic Bridging}

We define \emph{Semantic Bridging} as seeking features and knowledge to express requirements. It is about intention and specification layers.

For semantic bridging, we propose to jointly utilize semantic parsing and knowledge graph. On one side, semantic parsing converts natural language utterances to machine instructions~\cite{Lyu_2019}. It mines information from text data with techniques like named entity recognition and relation extraction.
On the other side, knowledge graphs use graph structures to organize semantic entities and their interrelations~\cite{Ji2021ASO}. It provides further flexibility by linking together with available commonsense or domain knowledge graphs.

The intention layer relies on a linguistic meaning theory called frame semantics~\cite{Fillmore2001FrameSF}.
A semantic frame is a conceptual structure of the involved participants and is used to describe events, scenes, etc.
It usually contains frame elements as semantic roles in the formal definition and lexical units that evoke other frames. That makes it very similar to a function module or class definition in programs.
This similarity inspires us to build a corpus oriented to program intentions, for example, based on \textsc{FrameNet} lexical database~\cite{Ruppenhofer2006FrameNetIE}.

The specification layer emphasizes the semantics information of code patterns, which is represented by its own description and the relations with others. The description could come from related text data, such as comments, API directives, and even identifiers. The relations could be meaning representation of text data~\cite{Xie2020APIMR}, or the logic representation of code data~\cite{Liu2021LearningbasedEO}. Compared with the intention layer, the specification layer differs in granularity, even though they both revolve around semantic entities and interrelations.

\subsection{Top-Down Pipeline}

In the framework, the top-down pipeline associates text code with code data through two transitions: the transition from intentions to specifications, and from specifications to realizations. Their medium data is respectively the semantic sketch and the code sketch~\cite{SolarLezama2006CombinatorialSF}.

In the transition from intentions to specifications, the semantic sketch purely cares about concepts and functions. Both semantic frame and semantic sketch take concepts as nodes and functions as edges.
However, compared with code sketches, their concepts could be more abstract while functions could be more complex.
Thereby, the transition aims to adopt alignment methods to eliminate the semantic gap between semantic sketch and code sketch, especially the granularity inconsistency. Other concerns are the details in the transition, such as issues on constraint and knowledge.

The transition medium from specifications to realizations is the code sketch. The code sketch is different from the code template, even though both are graphs taking code patterns as nodes. The code sketch takes semantic dependencies between code patterns as edges while the code template takes actual dependencies in samples as edges. Therefore, assuming we have a code sketch from the upstream, we could retrieve code templates for each code pattern in the given code sketch, then composite or generate code templates as candidates. Further, with manual inspection or intervention on elements and structures, the optimal code instance is determined.

\subsection{Scope of Application}

Due to the complexity and diversity in actual development, it is important to find suitable application scopes for the framework. Thereby, we considered some as preliminary targets.

One conceived application is assistive top-down developments for specific purposes or topics, including WebUI programming as mentioned.
Similarly, mobile application development is worth attention, since it follows a conventional standard to implement preset interactions. For example, an app can be recognized as a transition graph of activities, where each activity corresponds to a UI page and its functional code. The implementation of an activity is usually decoupled as layout code, functional code, and method call hierarchy~\cite{Chen2019StoryDroidAG}, and thus well fits the requirements of SPF.

In addition, to reveal the potentiality of SPF in associating intentions with realizations, automated construction of modular API dependency can be a topic. It elicits and reacts to potential development needs.
For example, a new web framework is at its early stage and wants a series of features, already implemented in similar frameworks but partially and scattly. A solution is to massively spawn and smoothly assemble the migrated implementations. Leveraging SPF, we refine the process and let it be effective and efficient.

% ..., code information of multiple layers in the semantic pyramid corresponds to different abstraction levels and also reveals the software hierarchy. It can be regarded as a multimodal representation of software~\cite{Weyssow2022BetterMT}. In this direction, we consider using code taxonomy and semantic pyramid for program representation.

%% file: tables/information.tex
\sisetup{table-format=2.2}
\rowcolors{2}{}{gray!10}
\begin{tabular}{
    llll
}

\hiderowcolors
\toprule

\textbf{Group} & \textbf{Type} & \textbf{Description} & \textbf{Example} \\

\midrule
\showrowcolors

\text{Attributes}
& \text{Constraint} & {restrictions on the applicable scope} & {conditions, inputs, outputs} \\
& \text{Element} & {discriminative implementation details} & {operators, numerics, strings} \\
& \text{Knowledge} & {conventions and usages in reusing code} & {API dependencies, frameworks} \\

\midrule

\text{Syntactics}
& \text{Relevance} & {connections among relevant information} & {sequence, syntax trees and graphs} \\

\midrule

\text{Semantics}
& \text{Concept} & {entities referred to and their interrelations} & {contexts, identifiers (objects, values)} \\
& \text{Function} & {logical functions to be actually executed} & {calculations, operations, exceptions} \\

\bottomrule

\end{tabular}

%% file: tables/taxonomy.tex
\sisetup{table-format=2.2}
\rowcolors{2}{}{gray!10}
\begin{tabular}{
    l ccc cc c
}

\hiderowcolors
\toprule

\multirow{2}[2]{*}{\textbf{Datatype}} & \multicolumn{3}{c}{\textbf{Attributes}} & \multicolumn{1}{c}{\textbf{Syntactics}} & \multicolumn{2}{c}{\textbf{Semantics}} \\
\cmidrule(lr){2-4} \cmidrule(lr){5-5} \cmidrule(lr){6-7}
& {\textbf{Constraint}} & {\textbf{Element}} & {\textbf{Knowledge}} & {\textbf{Relevance}} & {\textbf{Concept}} & {\textbf{Function}} \\

\midrule
\showrowcolors

\text{Code Pattern} & \xmark & \xmark & \cmark & ~ & ~ & ~ \\
\text{Code Template/Sketch} & \cmark & \xmark & \cmark & ~ & ~ & ~ \\
\text{Code Instance} & \cmark & \cmark & \cmark & ~ & ~ & ~ \\
\text{Code Property Graph} & ~ & ~ & ~ & \cmark & ~ & ~ \\
\midrule

\text{Semantic Entity} & ~ & ~ & ~ & ~ & \cmark & \xmark \\
\text{Semantic Pattern} & ~ & ~ & ~ & ~ & \xmark & \cmark \\
\text{Semantic Frame/Sketch} & ~ & ~ & ~ & ~ & \cmark & \cmark \\

\bottomrule

\end{tabular}

%% file: contents/3.roadmap.tex
\section{Roadmap}
\label{sec:roadmap}

To obtain experience in building the code taxonomy, and form an effective methodology, we divide the whole process into three steps.
First, we start with neural methods that learn from both code properties and data distributions, focusing on program representation and code generation.
Then, we build code taxonomies in limited application scopes and evaluate their effectiveness and drawbacks.
Finally, we seek decent models for a complete framework and increase the scale of the engineering to explore its applicability to real-world use.
The milestones to be achieved are as follows:

\begin{itemize}
    \item First, we would conduct a mining study on open source software distinguished by the purpose or topic. The object data is coding patterns, complemented by the information on context and API dependence.
    \item Further, we would seek solutions for semantic associations between program sketches and functional descriptions or API directives, but more focused on graph representations of multiple granularity or modality.
    \item To bridge intentions and specifications in a formal way, we plan on constructing a lexical database leveraging frame semantics and knowledge graphs. Besides, we connect it to available instances from relevant fields.
    \item To bridge specifications and realizations and promote code reusability, we plan on building a code corpus, where the data are templatized but composable subprograms, with the compatibility support of code variants.
    \item Once the code taxonomy is constructed, we would design retrieval-based neural models or inference engines to pair with this knowledge base. The framework should apply to other scenes where the semantic pyramid fits.
    \item Eventually, we plan to study the behavior and impact of the framework on building programs with a certain complexity, moreover, on building highly modular software or libraries that can satisfy actual development requirements.
\end{itemize}

%% file: contents/4.related_work.tex
\section{Related Work}
\label{sec:related_work}

For code taxonomy, we mainly refer to other practices on taxonomy for deep learning.
Meanwhile, for SP and SPF, we consider tasks connecting code data and text data as related work. Besides, we are aware of topics involved in implementing the framework, including code features, and topics about actual applications.

\paragraph{\bf{Taxonomy for Deep Learning}} When some tasks are hard to learn or evaluate, it is a strategy to build a taxonomy. For example, in software engineering, there is a taxonomy on build failures in continuous integration~\cite{Vassallo2017ATO}, a taxonomy of faults in deep learning systems~\cite{Humbatova2020TaxonomyOR}.
In other areas, taxonomy for deep learning even plays a considerable role. In natural language processing, \textsc{FrameNet} motivated the study on automatic semantic role labeling~\cite{Ruppenhofer2006FrameNetIE}.
In computer vision, \textsc{ImageNet}~\cite{Russakovsky2015ImageNetLS} is an image database for visual object recognition, it follows the \textsc{WordNet}~\cite{Miller1992WordNetAL} hierarchy, where each node is depicted by massive samples.

\paragraph{\bf{Code Retrieval, Generation and Comprehension}}
Sometimes, programmers want alternatives to current implementations, for example, \textsc{Aroma}~\cite{Luan2019AromaCR} recommends code snippets based on the similarity of structural features.
To search code snippets for the given query, \textsc{CODEnn} aligns the semantics of code data and text data by jointly modeling their representations in the same vector space~\cite{Gu2018DeepCS}. CSRS~\cite{Cheng2022CSRSCS} considers both relevance matching and semantic matching, while \textsc{Yogo}~\cite{Premtoon2020SemanticCS} recognizes different but mathematically-equivalent alternatives based on dataflow graphs and rewrite rules.
Inspired by machine translation, code generation translates descriptions to implementations, such as the pseudocode-to-code one by \textsc{SemanticScaffold}~\cite{Zhong2020SemanticSF}. \textsc{TRANX} introduced a transition system to generate formal meaning representations for code generation~\cite{Yin2018TRANXAT}. Code comprehension uses proper natural utterance to describe the given code data, which could help understand programs:
\textsc{Context2Name}~\cite{Bavishi2018Context2NameAD} generates meaningful variable names based on the usage contexts,
\textsc{AdaMo}~\cite{Gu2022AssembleFM} assembles foundation models to describe the functionality of code snippets, \textsc{CodeQA}~\cite{Liu2021CodeQAAQ} demonstrates the value of generating textual answers for the pairs of code snippet and various types of question.

\paragraph{\bf{Program Analysis and Reasoning}} By adopting changes to code data, the critical features can be stressed, such as \textsc{S4Eq}~\cite{Kommrusch2021SelfSupervisedLT}, which proves the equivalence by computing a verifiable sequence of rewrite rules, and \textsc{Perses}~\cite{Sun2018PersesSP}, which reduces programs to smaller ones in a syntax-guided way, and further help localizes bugs.
Also, program analysis can mine code features. For example, \textsc{Lupa}~\cite{Vlasova2022LupaAF} supports fine-grained analysis on the usage of programming language, \textsc{PyART}~\cite{He2021PyARTPA} extracts features from the context of the program point for API recommendation~\cite{Gu2016DeepAL}. Besides, code clone detection and code evolution promote code reuse~\cite{Zhang2022TheDA}.
In addition, neuro-symbolic approaches shown the reasoning capability in vision and natural language, and might work on code data, such as \textsc{Scallop}~\cite{Li2023ScallopAL}.

\paragraph{\bf{Assisted Programming and Development}}
There have proposed performant methods to assist programming, \textsc{CodeGen}~\cite{Nijkamp2022CodeGenAO} and \textsc{CodeT5}~\cite{Wang2021CodeT5IU} introduce large-scale language models to support common coding tasks, such as code autocompletion, code translation.
In contrast, assisted development focuses on the solution itself but not the programming process. For example, \textsc{AlphaCode}~\cite{li2022competition} applies sampling, filtering, clustering on program samples to generate human-level submissions for competitive programming challenges.
Some work is oriented to other tasks in the development, for example, \textsc{TransCoder} improves the application of back-translation for program translation~\cite{Lachaux2020UnsupervisedTO,Rozire2021LeveragingAU}, \textsc{SkCoder} reuses the code sketch, extracted from similar code, for reliable code generation~\cite{Li2023SkCoderAS}.

%% file: contents/5.conclusions.tex
\section{Conclusions}
\label{sec:conclusions}

To summarize, to overcome a few restrictions of neural methods, we suggest building a code taxonomy in light of code information. And then, we propose a semantic-aware framework, taking the semantic pyramid as its mental model, for top-down code generation.

SPF promotes code reusability through solid recognizability of code data and intense connectivity from expressibles to executables.
SPF indicates a novel framework of semantic awareness, especially when centered on the graph-form expression of code-related data. Its capability to support inspection or manipulation, requirement elicitation, and multimodal representation is beneficial.

As an exploratory work, top-down code generation is novel but faces challenges. For example, human-machine interactions could be very heavy in some cases where requirements are hard to capture or fulfill. Thus, we focus on realizing the proposed framework in limited scopes and seek continuous evolutions thereafter.